\documentclass[aps,preprint,amsmath,amssymb]{revtex4}
\usepackage{graphicx,axodraw}

\usepackage{epsfig,longtable}
\usepackage{makeidx}
\usepackage{varioref}
\usepackage{xr-hyper}
\usepackage{amsmath}
\usepackage[verbose,hypertexnames=false,bookmarksopenlevel=1,bookmarksopen,bookmarksnumbered,colorlinks,plainpages=false,linktocpage]{hyperref}

\begin{document}

\title
{\Large \bf Searching for doubly charged Higgs bosons in M\"{o}ller
scattering by resonance effects at linear $e^-e^-$  collider}

\author{
Chian-Shu~Chen\footnote{e-mail: chianshu@gmail.com},
Chao-Qiang~Geng\footnote{e-mail: geng@phys.nthu.edu.tw}, and
Dmitry~V.~Zhuridov\footnote{e-mail: zhuridov@phys.nthu.edu.tw}}
\affiliation{Department of Physics, National Tsing Hua University,
Hsinchu, Taiwan 300}

\date{\today}


\begin{abstract}
We discuss the  parity-violating left-right asymmetries (LRAs) in
M\"{o}ller scattering at the International Linear Collider (ILC)
induced by doubly charged Higgs bosons in models with $SU(2)_L$
triplet and singlet scalar bosons, which couple to the left- and
right-handed charged leptons, respectively. These bosons are
important in scenarios for the generation of the neutrino mass. We
demonstrate that the contributions to the LRAs from the triplet
and singlet bosons are opposite to each other. In particular,  we
show that the doubly charged Higgs boson from the singlet scalar
can be tested at the ILC by using the resonance effect.
\end{abstract}

\maketitle


The smallness of active neutrino masses is one of the most
challenging problems in particle physics.  The many interesting
scenarios proposed to explain it can be divided into the
categories of ones with and without the right-handed neutrinos
(RNs). For examples, the small neutrino masses can be generated
through the widely known seesaw mechanism~\cite{seesaw1,seesaw2}
with the RNs embedded in new physics at a high energy scale, such
as grand unified theories~\cite{GUTs} and extra
dimensions~\cite{ExtraD}, while models with extended Higgs sectors
can accomplish the goal without the RNs
\cite{Zee,twoloop,Higgs,Garayoa,CGN,Chen,CGZ}. The former is
difficult to test directly due to the fact that the RNs have no
standard-model (SM) interactions and the latter gets many
interesting features, such as the existence of the doubly charged
Higgs bosons $H^{\pm\pm}$ from the triplet (singlet) coupling to
the left- (right-) handed charged leptons to generate the neutrino
masses at tree~\cite{Higgs} (two-loop)
level~\cite{twoloop,CGN,Chen,CGZ} radiatively.

In this paper, we show that it is possible to distinguish the
models with these two types of $H^{\pm\pm}$ by studying M\"{o}ller
scattering~\cite{moller} of $e^-e^-\rightarrow e^-e^-$ at a linear
$e^-e^-$ collider, such as the International Linear Collider
(ILC)~\cite{ILC,ILC1}. With polarized initial electron beams, one
can define the parity-violating left-right asymmetry
(LRA)~\cite{AP} by
\begin{equation}
\label{AP}
A_P=\frac{{d\,\sigma_{LL}\over d\cos\theta}-{d\,\sigma_{RR}\over d\cos\theta}}
{{d\,\sigma_{LL}\over d\cos\theta}+{d\,\sigma_{RR}\over d\cos\theta}}\,,
\end{equation}
where  $LL$ and $RR$ denote the initial $e^-e^-$ polarizations,
and $\theta$ is the angle between the initial and final $ee$
beams. As the asymmetry can be measured with high accuracy  in the
process, it  provides an excellent opportunity to probe
new-physics effects. In particular, we will demonstrate that the
LRA is sensitive to the doubly charged Higgs bosons of
$H^{\pm\pm}$ at their mass poles.

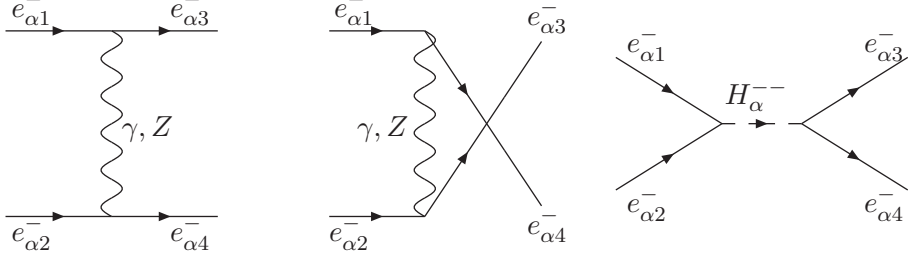
\begin{figure}[h]
\begin{picture}(330,100)(0,0)
\ArrowLine(0,80)(40,80)
\ArrowLine(40,80)(80,80)\Text(10,95)[t]{$e_{\alpha1}^-$}
\Text(70,95)[t]{$e_{\alpha3}^-$} \Photon(40,80)(40,10){4}{5}
\ArrowLine(0,10)(40,10)
\ArrowLine(40,10)(80,10)\Text(10,10)[t]{$e_{\alpha2}^-$}
\Text(70,10)[t]{$e_{\alpha4}^-$} \Text(44,43)[l]{$\gamma, Z$}

\ArrowLine(122,80)(158,80)\ArrowLine(158,80)(188,35)\Line(188,35)(202,14)
\Text(132,95)[t]{$e_{\alpha1}^-$}
\Text(205,92)[t]{$e_{\alpha3}^-$} \Photon(158,80)(158,10){4}{5}
\Text(133,43)[l]{$\gamma, Z$}
\ArrowLine(122,10)(158,10)\ArrowLine(158,10)(188,55)\Line(188,55)(202,76)
\Text(130,10)[t]{$e_{\alpha2}^-$}
\Text(205,13)[t]{$e_{\alpha4}^-$}

\ArrowLine(230,70)(270,45) \ArrowLine(230,20)(270,45)
\DashArrowLine(270,45)(300,45){4} \ArrowLine(300,45)(340,70)
\ArrowLine(300,45)(340,20) \Text(235,69)[lb]{$e_{\alpha1}^{-}$}
\Text(235,21)[lt]{$e_{\alpha2}^-$}
\Text(284,50)[b]{$H^{--}_\alpha$}
\Text(340,69)[rb]{$e_{\alpha3}^{-}$}
\Text(340,21)[rt]{$e_{\alpha4}^-$}
\end{picture}
\caption{Feynman diagrams of the M\"{o}ller scattering
for the contributions from $\gamma$, $Z$ and
$H_\alpha^{--}$.}
\label{F1}
\end{figure}

We start by considering a complex triplet scalar $H_{L}\equiv
T_{(2)}$ with the subscript denoting the hypercharge, which
couples to the $SU(2)_L$ lepton doublets ($L_{iL}$)~\cite{Higgs}:
\begin{eqnarray}\label{LLT}
    {\mathcal L}_{L}=g_{ij}\overline{L_{iL}^c}T L_{jL}+{\rm
    H.c.},
\end{eqnarray}
and a singlet scalar $H_R\equiv\Psi_{(4)}$ which couples to the
$SU(2)_L$ charged lepton singlets
($\ell_{iR}$)~\cite{CGN,Chen,CGZ}:
\begin{equation}\label{RR}
    {\mathcal L}_{R}=Y_{ij}\overline{\ell_{iR}^c}\ell_{jR}\Psi+{\rm
    H.c.},
\end{equation}
where $g_{ij}$ and $Y_{ij}$ are the coupling constants,
$i,j=e,\mu,\tau$, and $c$ stands for charge conjugation.
The contributions in the SM and those with $H_\alpha^{--}$
($\alpha=L,R$) to the M\"{o}ller scattering at tree level are
shown in Fig.~\ref{F1}.

In the SM \cite{PDG}, the  matrix elements for M\"{o}ller
scattering due to $\gamma$ and $Z$ are given by
\begin{eqnarray}
    M_{\gamma\alpha} &=& \frac{e^2}{t}(\bar e_3\gamma^\mu P_\alpha e_1)
    (\bar e_4\gamma^\mu P_\alpha e_2) -
    \frac{e^2}{u}(\bar e_4\gamma^\mu P_\alpha e_1)
    (\bar e_3\gamma^\mu P_\alpha e_2),
    \nonumber\\
    M_{Z\alpha} &=& \frac{C_\alpha^2}{\tilde t}(\bar e_3\gamma^\mu P_\alpha e_1)
    (\bar e_4\gamma^\mu P_\alpha e_2) -
    \frac{C_\alpha^2}{\tilde u}(\bar e_4\gamma^\mu P_\alpha e_1)
    (\bar e_3\gamma^\mu P_\alpha e_2),
\end{eqnarray}
where $P_{L(R)}={1\over 2}(1\mp \gamma_5)$, $a=t, u$ is for the
Mandelstam variables, $\tilde a=a-M_Z^2$
and
\begin{eqnarray}
    C_L=\frac{g(1-2\sin^2\theta_W)}{2\cos\theta_W}, \quad
    C_R=\frac{g\sin^2\theta_W}{\cos\theta_W}.
\end{eqnarray}
The non-SM interactions  in Eqs. (\ref{LLT}) and (\ref{RR}) can be
combined into a single form with the doubly charged Higgs bosons
$H_\alpha^{++}$ and $i=j=e$ as follows:
\begin{eqnarray}
    {\mathcal L}_{H_\alpha} = Y_{\alpha ee}\overline{e^c}P_\alpha eH_\alpha^{++}
    + {\rm H.c.}\,,
\end{eqnarray}
which leads to
\begin{eqnarray}
    M_{H_\alpha} = |Y_{\alpha ee}|^2(\bar e_2^cP_\alpha e_1)\frac{1}{s-M_{H_\alpha}^2-
    iM_{H_\alpha}\Gamma_{H_\alpha}}(\bar e_4P_\alpha e_3^c)\,,
\end{eqnarray}
with the total widths  \cite{Chen}
\begin{eqnarray}\label{widthL}
\Gamma_{T}&=&3\left[\Gamma_L(\ell_{i}^{\pm}\ell_{i}^{\pm})+\Gamma_L(\ell_{i}^{\pm}\ell_{j}^{\pm})_{i\neq
j}\right]+ \Gamma_L(W^{\pm}W^{\pm})+ \Gamma_L(W^{\pm}P^{\pm})+
\Gamma_L(W^{\pm}W^{\pm}T^0_a),\nonumber\\
\Gamma_{\Psi}&=&3\left[\Gamma_R(\ell_{i}^{\pm}\ell_{i}^{\pm})+\Gamma_R(\ell_{i}^{\pm}\ell_{j}^{\pm})_{i\neq
j}\right],
\end{eqnarray}
where $Y_{\alpha ee}$ are new coupling constants related to
$g_{ee}$ and $Y_{ee}$, and
\begin{eqnarray}\label{gamma}
\Gamma_\alpha(\ell_{i}^{\pm}\ell_{j}^{\pm})&=&(2-\delta_{ij})\frac{|Y_{\alpha ij}|^2}
{16\pi}M_{H_\alpha},
\nonumber\\
\Gamma_L(W^{\pm}W^{\pm})&=&\frac{g^4v_T^2}{16\pi M_T}
\sqrt{1-\frac{4M_W^2}{M^2_{T}}}\left(3-\frac{M_{T}^2}{M_W^2}+\frac{M^2_{T}}{4M_W^2}\right),
\nonumber\\
\Gamma_L(W^{\pm}P^{\pm})&=&\frac{g^2M^3_{T}}{16\pi
M_W^2}\lambda^{3/2}
\left(1,\frac{M_W^2}{M_{T}^2},\frac{M_P^2}{M_{T}^2}\right),
\end{eqnarray}
with $\lambda(x,y,z)=x^2+y^2+z^2-2xy-2xz-2yz$ and $P^{\pm}$ and $T_a^0$ being the single-charged and neutral components of the Higgs scalars,
respectively.
Note that the subscript  $a$ in $T_a^0$ denotes the
mass eigenstate, which is a mixture of the doublet and triplet bosons.
The three-body decay modes in Eq.~(\ref{widthL}) are expected to be relatively
suppressed by the phase space compared to the two-body ones.
Note that we have assumed the particular model to include either
one of the two new interactions or both. Since the triplet scalar
contains the SM quantum numbers, the gauge interactions provide
more decay channels for its doubly charged component.

In the scenario with the interaction in Eq.~(\ref{LLT}), the
neutrino masses are proportional to $g_{ij}v_T$, where $v_T$
denotes the VEV of
$T$. However, from the neutrino oscillation data and cosmological
experiments there are restricting bounds on the neutrino masses,
given by~\cite{nu_mass}
\begin{eqnarray}\label{bound}
m_{\nu} \sim g_{ij}v_T \lesssim 0.1\ {\rm eV}\,.
\end{eqnarray}
 On the other hand, the precision data of $\rho=1.0002^{+0.0007}_{-0.0004}$ \cite{PDG}
 result in a  limit of
 $v_T$ that is less than $ 4.41$~GeV \cite{Chen}.
We shall use $v_T$~= 4~GeV in our numerical calculations.

The parity-violating LRA in Eq. (\ref{AP}) can now be rewritten
as
\begin{eqnarray}\label{AP1}
    A_{P} = \frac{\sum\limits_{spin}|M_{L}|^2-\sum\limits_{spin}|M_{R}|^2}
    {\sum\limits_{spin}|M_{L}|^2+\sum\limits_{spin}|M_{R}|^2},
\end{eqnarray}
where
\begin{eqnarray}\label{Ma}
    \sum\limits_{spin}|M_\alpha|^2 =
    \sum\limits_{spin}|M_{\alpha}|^2_{SM} +
    \sum\limits_{spin}|M_{H\alpha}|^2\,.
\end{eqnarray}
In Eq. (\ref{Ma}),  the SM contributions are
\begin{eqnarray}
\sum\limits_{spin}|M_{\alpha}|^2_{SM} =
\sum\limits_{spin}|M_{\gamma\alpha}|^2 +
\sum\limits_{spin}|M_{Z\alpha}|^2 +
\left(\sum\limits_{spin}M_{\gamma\alpha}M_{Z\alpha}^\dag +
\sum\limits_{spin}M_{Z\alpha}M_{\gamma\alpha}^\dag\right),
\end{eqnarray}
where
\begin{eqnarray}
\nonumber
    \sum\limits_{spin}|M_{\gamma\alpha}|^2 = \left[ 2e^2\frac{s(t+u)}{tu}
    \right]^2, \quad\quad \sum\limits_{spin}|M_{Z\alpha}|^2 =
    \left[ 2C_\alpha\frac{s(\tilde t+\tilde u)}{\tilde t\tilde u}
    \right]^2,
\end{eqnarray}
\begin{eqnarray}
    \sum\limits_{spin}M_{\gamma\alpha}M_{Z\alpha}^\dag = \sum\limits_{spin}M_{Z\alpha}
    M_{\gamma\alpha}^\dag = 4e^2C_\alpha s^2\left[ \frac{1}{t\tilde
    t}+\frac{1}{t\tilde u}+\frac{1}{u\tilde t}+\frac{1}{u\tilde u}
    \right],
\end{eqnarray}
and the non-SM ones due to the doubly charged Higgs bosons are
\begin{eqnarray}
    \sum\limits_{spin}|M_{H\alpha}|^2 = |Y_{\alpha ee}|^4
    \frac{s^2}{(s-M_{H_\alpha}^2)^2+M_{H_\alpha}^2\Gamma_{H_\alpha}^2}.
\end{eqnarray}
An interesting point here is that the interference between the SM
and the new interactions, i.e., the vector--scalar interference
term, vanishes in this calculation.

\begin{figure}[ht]
  \centering
   \includegraphics*[width=2.8in]{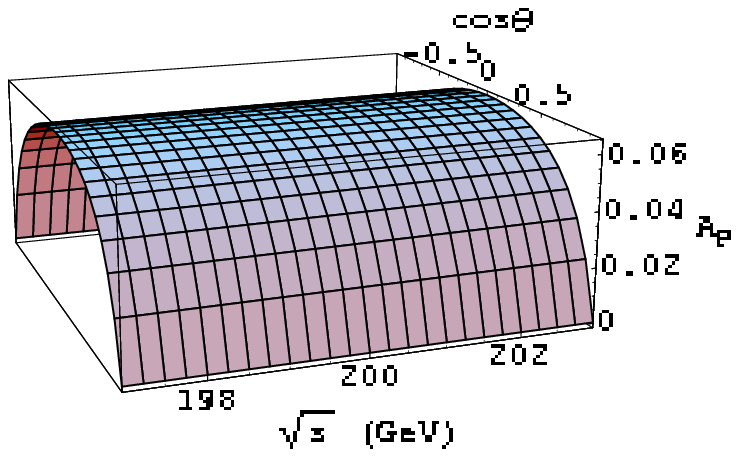}
     \includegraphics*[width=2.8in]{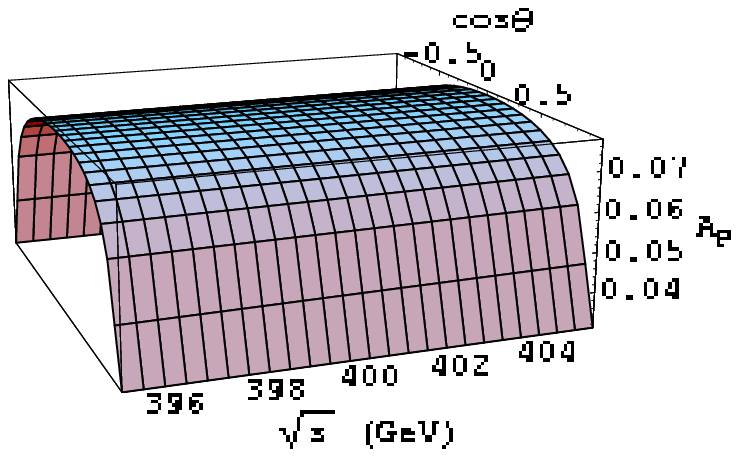}
      \includegraphics*[width=2.8in]{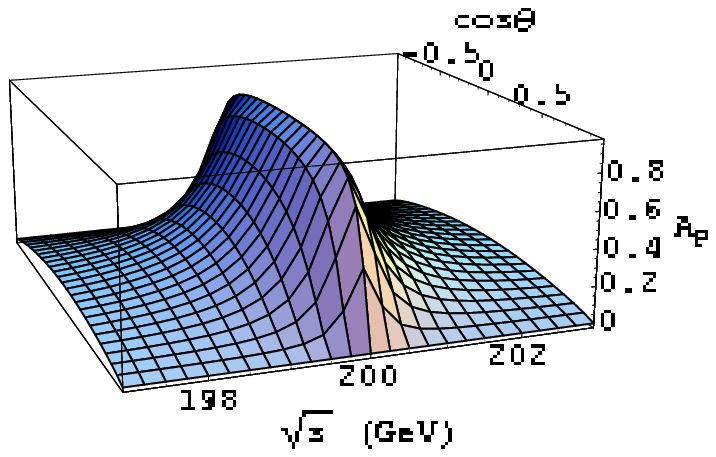}
     \includegraphics*[width=2.8in]{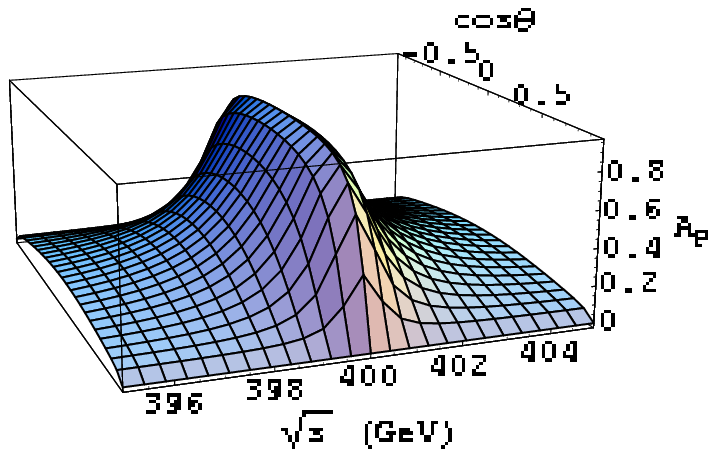}
     \includegraphics*[width=2.8in]{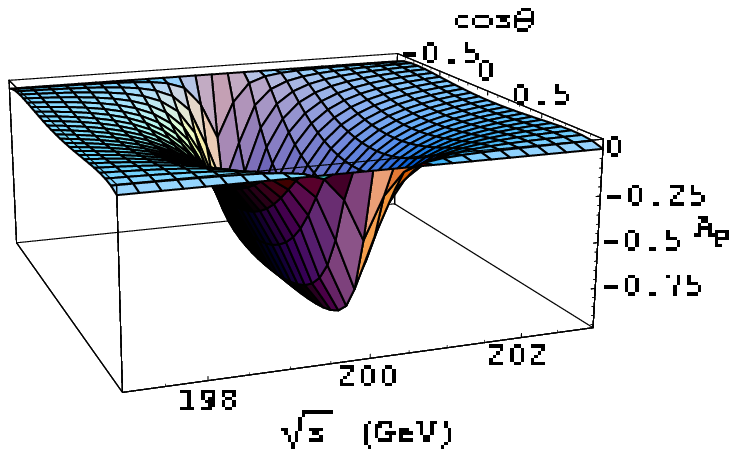}
     \includegraphics*[width=2.8in]{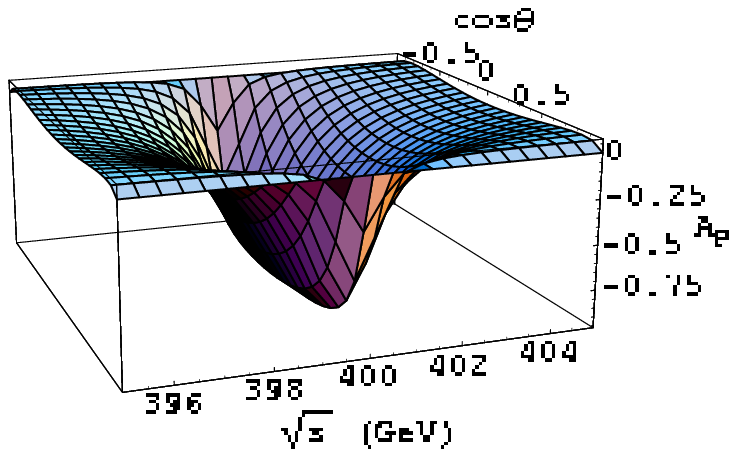}
  \caption{
$A_P$ versus $\cos\theta$ for the energies around $\sqrt{s}$~=
200~GeV ({\it left}) and 400~GeV ({\it right}) in the SM ({\it
upper}) and the models with the triplet ({\it middle}) and singlet
({\it lower}) scalar bosons, where $M_{H_\alpha}=200$~GeV ({\it
left}) and 400~GeV ({\it right}), respectively.}
\label{F2}
\end{figure}
The results for $A_P$ versus $\cos\theta$ are presented in
Fig.~\ref{F2} for the SM and the models with the triplet and
singlet bosons in the interactions of Eqs. (\ref{LLT}) and
(\ref{RR}), respectively. In the figures (middle and lower),
 the resonance energies for the non-SM cases are
considered, the doubly charged Higgs boson masses  are taken to be $M_{H\alpha}$~=
200 ({\it left}) and 400~GeV ({\it right}) with the couplings
$|Y_{\alpha ee}|$~= 0.1 close to the upper bounds~\cite{Chen}, and  the
values of $\Gamma_{H\alpha}$ are
given in Table~\ref{T1}.
\begin{table}[h]
\caption{The values of $\Gamma_{H\alpha}$ for the various values
of $M_{H\alpha}$.} \label{T1}
\begin{tabular}{|c|c||c|c|c|}
  \hline
  $|Y_{\alpha ee}|$ & $M_{H\alpha}$, GeV & $\Gamma_T$, GeV & $\Gamma_\Psi$, GeV \\
  \hline
  \hline
  0.1 & 200 &  0.359 & 0.358 \\
  0.1 & 400 &  0.736 & 0.716 \\
  0.05& 200 &  0.091 & 0.090 \\
  0.02& 200 &  0.015 & 0.014 \\
  0.005& 200 &  0.0020 & 0.00090 \\
  0.002& 200 &  0.0012 & 0.00014
  \\\hline
\end{tabular}
\end{table}
The current lower bound on the doubly charged scalar mass set by
the direct search at the Tevatron in Fermilab is 136~GeV
\cite{Tevatron}. Since the doubly charged Higgs bosons in Eqs.
(\ref{LLT}) and (\ref{RR}) couple to the different helicity states
of the electron, the effects lead to an opposite sign of $A_P$.
Thus, one can distinguish the two types of the interactions based
on the sign of $A_P$ around the resonance.

\begin{figure}[ht]
  \centering
    \includegraphics*[width=2.8in]{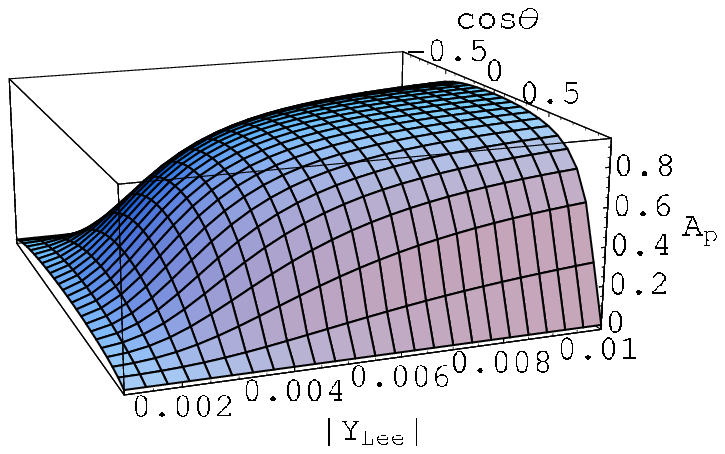}
     \includegraphics*[width=2.8in]{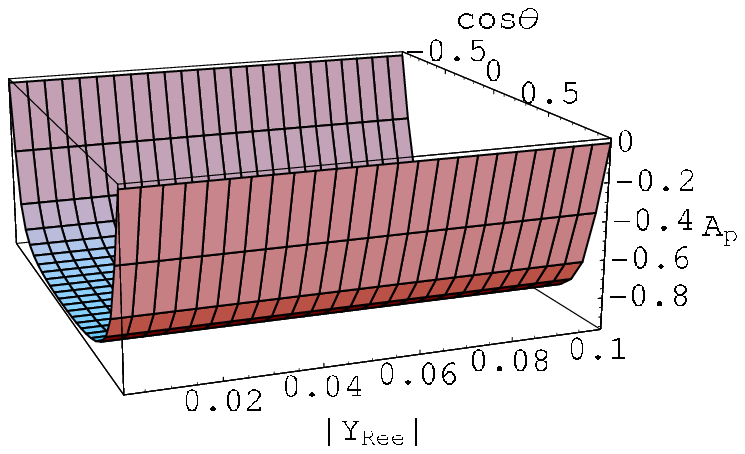}
     \includegraphics*[width=2.8in]{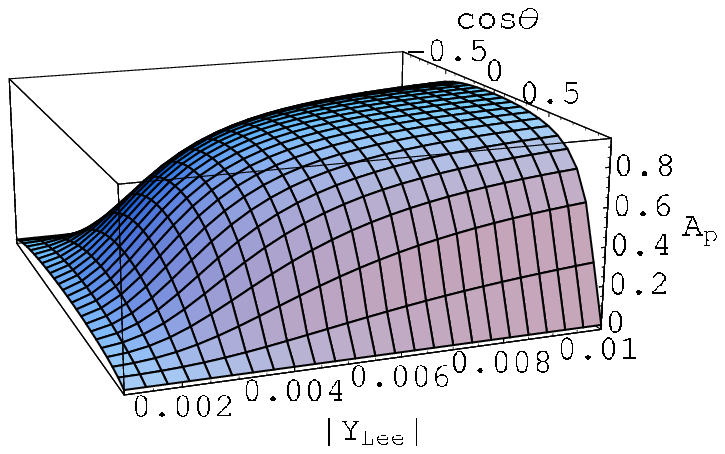}
     \includegraphics*[width=2.8in]{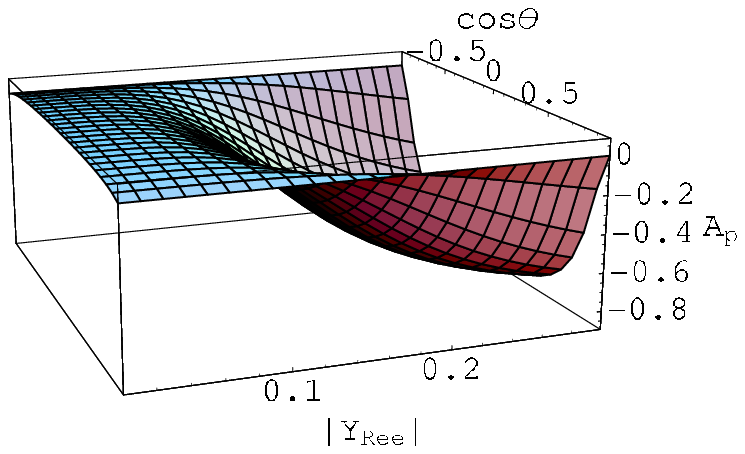}
  \caption{$A_p$ versus $\cos\theta$ and $|Y_{\alpha ee}|$
  in the resonance point $\sqrt{s}$~= $M_{H\alpha}$~= 200~GeV  ({\it
  upper}),
  the near one of $\sqrt{s}$~= 202~GeV ({\it lower})
  with
  the doubly charged Higgs bosons from the triplet ({\it left}) and singlet ({\it right}) scalars.}
  \label{F3}
\end{figure}

It should be stressed that these effects are strongly dependent on
the values of the couplings $Y_{\alpha ee}$, since the non-SM
effects are proportional to $|Y_{\alpha ee}|^4$, and the widths
$\Gamma_{\alpha}(e^\pm e^\pm)$ are proportional to  $|Y_{\alpha
ee}|^2$. This is shown for the resonance point of $\sqrt{s}$~=
$M_{H\alpha}$~= 200~GeV and the near one of $\sqrt{s}$~= 202~GeV
in Fig.~\ref{F3}.
By taking into account the strong bounds in Eq.~(\ref{bound}) on
the couplings $g_{ij}\sim Y_{Lij}$ in Eq.~(15), one may conclude
that there is no chance to observe the LRA due to the  interaction
in Eq. (\ref{LLT})  for the triplet scalar in the near future. On
the other hand, the constraints on the interaction in Eq.
(\ref{RR}) for the singlet scalar are much relaxed~\cite{Chen},
providing a good opportunity to detect the effects at the ILC in
its $e^-e^-$ mode~\cite{ee_mode}.
As illustrated in Fig.~\ref{F3}, in the case of $\Psi$ there is
not precisely a dependence on $|Y_{\alpha ee}|$ in the resonance
point of $\sqrt{s}=M_{H\alpha}$. This is because the factor
$|Y_{\alpha ee}|^4$ in the numerator and the factor $|Y_{\alpha
ee}|^2$ in the width
cancel each other.

We now consider the models that include the complex triplet and singlet bosons,
simultaneously. In this kind of models, the two doubly charged
Higgs bosons mix to each other with the mixing angle $\delta$ to form the
mass eigenstates $P_{1,\,2}$:
\begin{eqnarray}
\left(\begin{array}{c}P_1^{\pm\pm}
\\P_2^{\pm\pm}\end{array}\right) =
\left(\begin{array}{cc}\cos{\delta} & \sin{\delta} \\-\sin{\delta}
& \cos{\delta}\end{array}\right)\left(\begin{array}{c}T^{\pm\pm}
\\\Psi^{\pm\pm}\end{array}\right).
\end{eqnarray}
Therefore, the interactions in Eqs. (\ref{LLT}) and (\ref{RR}) can be rewritten  as
\begin{eqnarray}
{\mathcal L}_{LM} &=& g_{ij}\overline{\ell_{Li}^c}\ell_{Lj}
(\cos{\delta}P_1 - \sin{\delta}P_2)+{\rm H.c.},\nonumber \\
{\mathcal L}_{RM} &=&
Y_{ij}\overline{\ell_{Ri}^c}\ell_{Rj}(\sin{\delta}P_1 +
\cos{\delta}P_2)+{\rm H.c.}\,,
\end{eqnarray}
which become the pure left- and right-handed interactions for the
limit $\delta\rightarrow0$, respectively. Here, we will
concentrate on the lighter massive state $P_1$. The considered
models share both features of the models with only the interaction
in Eq. (\ref{LLT}) or (\ref{RR}). Due to the strong constraint on
the couplings in Eq. (\ref{LLT}), these models can give a
significant contribution only to $d\sigma_{RR}$ in Eq. (\ref{AP}),
which tends to flip the sign of  $A_P$. On the other hand all the
decay channels in Eq.~(\ref{widthL}) are permitted for $P_1$ due
to the mixing in the doubly charged Higgs sector.

\begin{figure}[ht]
\centering
\includegraphics*[width=2.8in]{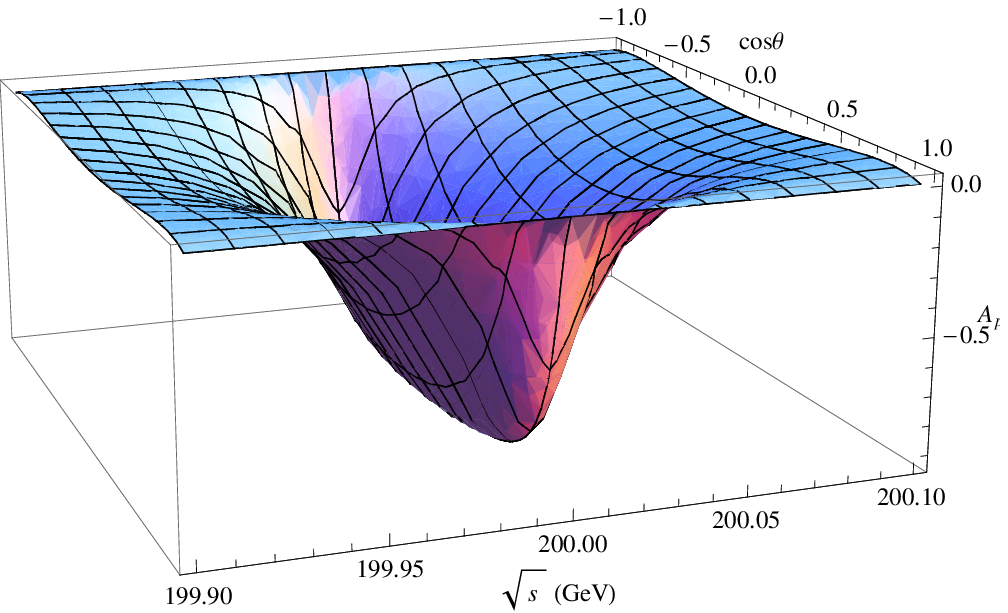}
\includegraphics*[width=2.8in]{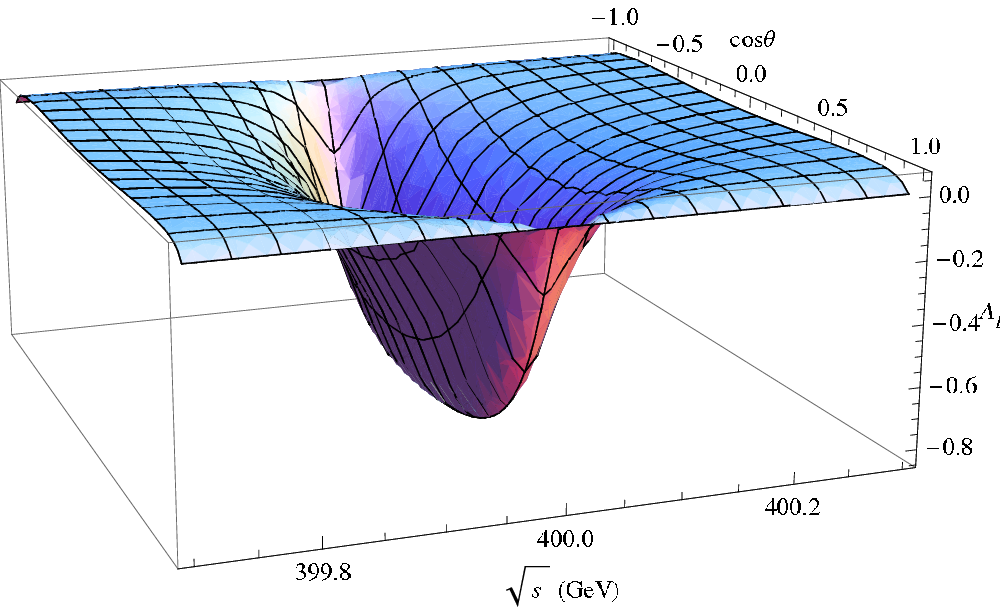}
\caption{$A_P$ in the general model at the resonance points of
200~GeV ({\it left}) and 400~GeV ({\it right}) with small mixing 
and large mixing ({\it see separate files alongside the text}) angles.}
\label{F4}
\end{figure}

\begin{figure}[ht]
\centering
\includegraphics*[width=2.8in]{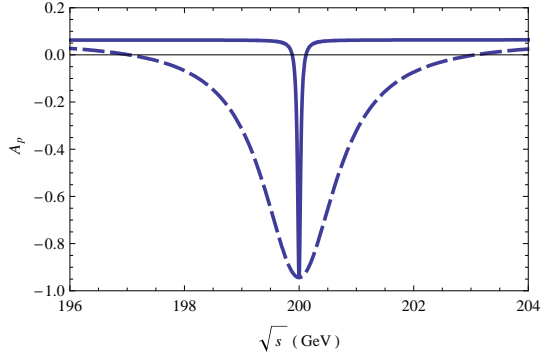}
\caption{$A_P$ versus $\sqrt{s}$ for $M_{P_1}=200$~GeV for pure
singlet and small mixing (coincided solid lines),
 large mixing (dashed line).}
 \label{F5}
\end{figure}

In the model proposed in Refs.~\cite{CGN,Chen}, as there is no
tree-level triplet Yukawa interaction in Eq. (\ref{LLT}), only the
singlet interaction in Eq. (\ref{RR}) is allowed. In terms of the
triplet gauge couplings and singlet Yukawa couplings the decay
channels $P_1^{\pm\pm}\rightarrow \ell_{iR}^{\pm}\ell_{jR}^{\pm}$,
$P_1^{\pm\pm} \rightarrow W^{\pm}W^{\pm}$, $P_1^{\pm\pm}
\rightarrow W^{\pm}P^{\pm}$ and $P_1^{\pm\pm}\rightarrow
W^{\pm}W^{\pm}T^0_a$ are permitted. We remark that the widths of
these decays have an additional mixing-angle dependence compared
with those in Eq. (\ref{gamma})~\cite{Chen}. As  shown in
Fig.~\ref{F4}, the deviations from the SM, allowed by the present
experimental bounds,
can be observed at the ILC.
In our discussion, we have taken the  upper bound of $|Y_{ee}|=0.2$~\cite{Chen} and the
small (large) mixing of $\sin\delta=0.1$ (0.5). The  values of
$\Gamma_{P_1}$ are given in Table~\ref{T2}.
\begin{table}[htb]
\begin{center}
\caption{The values of $\Gamma_{P_1}$ for $M_{P_1}$=200 and 400
GeV and $\sin\delta$=0.1 and 0.5.} \label{T2}
\begin{tabular}{|c|c||c|c|}
  \hline
  $\sin\delta$ & $M_{P_1}$, GeV & $\Gamma_{P_1}$, GeV \\
  \hline
  \hline
  0.1 & 200 &  0.015 \\
      & 400 &  0.048 \\
  \hline
  0.5 & 200 &  0.359 \\
      & 400 &  0.732 \\
  \hline
\end{tabular}
\end{center}
\end{table}
The valleys in Fig.~\ref{F4} are much wider in the case with the
large mixing than those with the small mixing. This can help us to
clarify the mixing between the doubly charged scalars.
For example, Fig.~\ref{F5} shows that it is easy to separate the
model with the large mixing by using the $A_P$ versus $\sqrt{s}$
dependence, while models with the small mixing and pure singlet
(no mixing) are practically undistinguishable. Measuring of the
width and height of the resonance  could determine $Y_{\alpha ee}$
and $\sin\delta$, which are relevant to the neutrino mixings.

The relative deviations of $A_P$ from the SM values of $A_P^{SM}$,
defined by $\alpha_P=(A_P^{SM}-A_P)/A_P^{SM}$ are given in
Table~\ref{T3} for various  resonance energies $\sqrt{s}$ at the
maximum points of $\cos\theta=0$ with $M_{P_1}= 200$ and 400~GeV
and  $\sin\delta= 0.1$ and 0.5. One can see that measuring
$\alpha_P$ with 10\% (1\%) accuracy and fixing the energy with 1\%
accuracy or less allows one to observe the resonance effects in
the large (small) mixing. This accuracy can be achieved with
realistic polarizations for the initial electrons, since the beam
energy and polarizations are expected to be stable and measurable
at a level of about 0.1\%~\cite{ILC}. Taking the electron
longitudinal polarizations of 0.8 and assuming their systematic
uncertainties of 0.5\%~\cite{ee_mode}, we estimate that the
corresponding uncertainty in measuring $A_P$ is about 1\%. At the
resonance, the non-standard contributions to the LRA are dominant.
The dependence on the mixing angle in Eq.~(\ref{AP1}) almost
disappears since the terms with $Y_{Lij}$ are small. For this
reason, $\alpha_P$ is about the same for the different mixings at
the resonance.

The integrated deviations $\tilde\alpha_P=(\tilde A_P^{SM}-\tilde A_P)/\tilde A_P^{SM}$, with
\begin{equation}
\label{APintegrated}
\tilde A_P=\frac{\sigma_{LL}-\sigma_{RR}}
{\sigma_{LL}+\sigma_{RR}}\,,
\end{equation}
are given in the last three columns in Table~\ref{T3} for the two types of mixing and
the pure singlet ($\sin\theta=1$). In Eq.~(\ref{APintegrated}), we have integrated
the differential cross sections 
over $\cos\theta$ in the range $\cos1^\circ$ to $\cos179^\circ$,
since coverage down to $\sim1^\circ$ from the beam axis is
expected at ILC~\cite{ILC1}.  From Table~\ref{T3}, we see that
$\tilde\alpha_P$
decreases with increasing $M_{P_1}$ and slightly depends on the mixing angle.

\begin{table}[htb]
\begin{center}
\caption{ Relative deviations $\alpha_P$ of $A_P$  at
$\cos\theta=0$ for the various $M_{P_1}$ and mixings $s_\delta \equiv\sin\delta$ in and around
the resonance energies. The integrated deviations $\tilde\alpha_P$ of $\tilde A_P$ are also given.}\label{T3}
\begin{tabular}{|c|c||c|c||c|c|c|}
  \hline
    $M_{P_1}$, GeV & $\sqrt{s}$, GeV & $\alpha_P$, $s_\delta=0.1$ & $\alpha_P$, $s_\delta=0.5$ & $\tilde\alpha_P$, $s_\delta=0.1$ & $\tilde\alpha_P$, $s_\delta=0.5$ & $\tilde\alpha_P$, $s_\delta=1$ \\
  \hline
  \hline
   200& 200         & 0.47  & 0.47 & 38.0 & 41.4 & 41.6 \\
      & 200 $\pm$ 1 & 0.015 & 0.46 &&&\\
      & 200 $\pm$ 2 & 0.004 & 0.41 &&&\\
      & 200 $\pm$ 4 & 0.001 & 0.28 &&&\\
  \hline
   400& 400         & 0.33  & 0.35 & 5.5 & 13.5 & 14.1 \\
      & 400 $\pm$ 1 & 0.039 & 0.35 &&&\\
      & 400 $\pm$ 2 & 0.012 & 0.34 &&&\\
      & 400 $\pm$ 4 & 0.003 & 0.30 &&&\\
  \hline
\end{tabular}
\end{center}
\end{table}

In conclusion, we have investigated the contributions to the LRAs
from the doubly charged Higgs bosons in models with triplet and
singlet scalar bosons, which couple to the electrons with
different chiralities in M\"{o}ller scattering, respectively. We
have found that it is easy to extract the properties of the models
with $H^{\pm\pm}$ interacting with $e_R$ from the measurements of
$A_P$ around the resonance point at the ILC, whereas it is hard to
observe the effects of the models with $H^{\pm\pm}$ coupled only
to $e_L$. As the models have different mechanisms for the
generation of the neutrino mass, future searches for $A_P$ at the
ILC are important for us to understand the neutrino physics.
Finally, we remark that the $H^{\pm\pm}$ masses for the resonance
studies at the ILC can be determined by the searches at the Large
Hadron Collider (LHC), such as the dilepton signatures at the LHC
shown in Ref. \cite{CGZ}.

\section*{Acknowledgements} This work is financially supported by
the National Science Council of Republic of China under the
contract \#: NSC-95-2112-M-007-059-MY3. One of us (DVZ) would like
to thank TILK08 for hospitality in Sendai where this research was
presented.

\end{document}